\documentclass[11,draftclsnofoot, peerreview, onecolumn]{IEEEtran}
\usepackage{cite}
\usepackage{bm}

\ifCLASSINFOpdf
   \usepackage[pdftex]{graphicx}
\else
   \usepackage[dvips]{graphicx}
\fi
\usepackage{array}
\newcolumntype{C}{>{\centering\arraybackslash}p{2cm}}

\usepackage{bm}

\usepackage[cmex10]{amsmath}

\usepackage{mdwmath}
\usepackage{subfigure}
\usepackage{verbatim}
\begin{document}
%
\title{On the Full Column-Rank Condition of the Channel Estimation Matrix in Doubly-Selective MIMO-OFDM Systems}

\maketitle

\begin{abstract}
Recently, this journal has published a paper which dealt with basis expansion model (BEM) based least-squares (LS) channel estimation in doubly-selective orthogonal frequency-division multiplexing (DS-OFDM) systems. The least-squares channel estimator computes the pseudo-inverse of a channel estimation matrix. For the existence of the pseudo-inverse, it is necessary that the channel estimation matrix be of full column rank. In this paper, we investigate the conditions that need to be satisfied that ensures the full column-rank condition of the channel estimation matrix. In particular, we derive conditions that the BEM and pilot pattern designs should satisfy to ensure that the channel estimation matrix is of full column rank. We explore the polynomial BEM (P-BEM), complex exponential BEM (CE-BEM), Slepian BEM (S-BEM) and generalized complex exponential BEM (GCE-BEM). We present one possible way to design the pilot patterns which satisfy the full column-rank conditions. Furthermore, the proposed method is extended to the case of multiple-input multiple-output (MIMO) DS-OFDM systems as well. Examples of pilot pattern designs are presented, that ensure the channel estimation matrix is of full column rank for a large DS-MIMO-OFDM system with as many as six transmitters, six receivers and 1024 subcarriers.


\end{abstract}


%
\IEEEpeerreviewmaketitle
\section{Introduction}
\PARstart{A}{} basis expansion model (BEM) based \cite{basis_PIEEE_98} least squares (LS) channel estimation scheme for doubly-selective orthogonal frequency-division multiplexing (DS-OFDM) system was proposed by Tang et al in \cite{pilot_JSP_07}. For good performance, the channel estimation matrix $\boldsymbol{\mathcal P}$ needs to be of full column rank \cite[pp. 2232]{pilot_JSP_07}. However, the full column-rank condition was not addressed in \cite[pp. 2232]{pilot_JSP_07}. The purpose of this paper is to derive the conditions on the BEM and pilot patterns such that the $\boldsymbol{\mathcal P}$ is of full column rank. Four BEMs are analyzed, namely, polynomial BEM (P-BEM) \cite{pilot_JSP_07}, complex exponential BEM (CE-BEM) \cite{pilot_JSP_07}, Slepian BEM (S-BEM) \cite{slepian_spl}-\cite{slepian_jsp} and generalized complex exponential BEM (GCE-BEM) \cite{pilot_JSP_07}.  

A BEM-based channel estimation scheme for doubly-selective multiple-input multiple-output OFDM (DS-MIMO-OFDM) system was proposed by Song and Lim in \cite{comments}. The methods of \cite{pilot_JSP_07} and \cite{comments} are similar in spirit although \cite{pilot_JSP_07} did not deal with MIMO systems. Full column-rank conditions on the channel estimation matrix were not addressed in \cite{comments}. In this paper, the full column-rank conditions is extended to doubly-selective MIMO OFDM (DS-MIMO-OFDM) systems as well. {\it One possible way to design pilot patterns, that satisfy the full column-rank conditions, is presented in this paper. The pilot-pattern design examples are applicable to DS-MIMO-OFDM systems as well. To the best of the author's knowledge these pilot-pattern design examples have not been presented before in the literature.}

Tang et al authored a subsequent paper \cite{pilot_SPAWC_07} which deals with the full column-rank condition of $\boldsymbol{\mathcal P}$. The context and scope of the results presented in this paper differ with those presented in \cite{pilot_SPAWC_07}. The key differences between the analysis in this paper and \cite{pilot_SPAWC_07} can be summarized as follows:
\begin{itemize}
	\item The work of \cite{pilot_SPAWC_07} assumed a frequency-domain Kronecker delta (FDKD) pilot pattern. The analysis in this paper is not restricted to any particular pilot pattern. The FDKD pilot pattern is treated as a special case in this paper.
	\item The work in \cite{pilot_SPAWC_07} showed that the FDKD pilot pattern ensures that $\boldsymbol{\mathcal P}$ is of full column rank for CE-BEM and P-BEM. In this paper, we derive general conditions that a BEM should satisfy for $\boldsymbol{\mathcal P}$ to be of full column rank when any pilot pattern is used. Four BEMs are investigated, namely, P-BEM, CE-BEM, S-BEM and GCE-BEM. 
	\item We also present a detailed approach to pilot-pattern design that enables $\boldsymbol{\mathcal P}$ to be of full column rank. This design example is extended to DS-MIMO-OFDM systems as well. Furthermore, our analysis show that the design of pilot patterns can be independent of the conditions that a BEM should satisfy.
	\item This paper presents full column-rank conditions for DS-MIMO-OFDM systems. The work in \cite{pilot_SPAWC_07} is limited to DS-OFDM systems.
	\item Where possible, we also show that some of the results derived in this paper is in agreement with the ones derived in  \cite{pilot_SPAWC_07}.
	\end{itemize}
Cao and Li presented some full column-rank conditions of channel estimation matrix for DS-MIMO-OFDM systems \cite[{\it Lemma } 1]{commentscomments}. We show that some of the results arrived at in this paper are in agreement with \cite[{\it Lemma } 1]{commentscomments}.

The paper is organized as follows. The BEM-based OFDM system model is presented in Section II. This forms the basis for the full column-rank conditions discussed in Section III. Pilot-pattern design examples for DS-OFDM and DS-MIMO-OFDM systems are presented in Section IV, while Section V deals with miscellaneous conditions and consensus of results with prior art. Simulation studies are presented in Section VI and Section VII concludes the paper.

\subsection{Notation}
A boldface large and small letter mean a matrix and a vector, respectively. The quantity $\left\langle x\right\rangle$ denotes $x$ mod $N$ ($N$ is the number of subcarriers in the OFDM system) while $\left\langle x\right\rangle_a$ denotes $x$ mod $a$.  $(.)^H$, $(.)^T$ and $*$ denote Hermitian, transpose and conjugate, respectively. The $(p,q)$th and $p$th elements of ${\bf A}$ and ${\bf x}$ are denoted by ${\bf A}(p,q)$ and ${\bf x}(p)$, respectively. The element-wise product of two matrices, called as Hadamard product, is denoted by $\odot$, i.e., the $(i,j)$th element of ${\bf A}\odot{\bf B}$ is ${\bf A}(i,j){\bf B}(i,j)$. ${\bf I}_N$ represents an $N\negmedspace\times\negmedspace N$ identity matrix. Where possible, we use the Matlab notation with the exception being vector and matrix indices, which start from zero.  The notation ${\bf B} = e^{a{\bf Q}}$ means that $\bf B$ is a matrix whose dimension is the same as that of $\bf Q$ and ${\bf B}(i,j)=e^{a{\bf Q}(i,j)}$. We will use ${\bf A}(:,m\negmedspace:\negmedspace n)({\bf A}(m\negmedspace:\negmedspace n,:))$ to extract the submatrix from column (row) $m$ to column (row) $n$, ${\bf A}({\bf r}, {\bf c})$ to extract a submatrix within ${\bf A}$ defined by the index-vector of desired rows in ${\bf r}$, and the index-vector of desired columns in $\bf c$. The vector ${\bf  x}(m\negmedspace:\negmedspace n)$ extracts entries from  $m$ to $n$ and ${\bf x}({\bf r})$ extracts entries denoted by the index-vector $\bf r$. The $N\negmedspace\times\negmedspace N$ FFT matrix $\bf F$ is defined as ${\bf F}(m,n)\negmedspace=\negmedspace e^{-\frac{j2\pi mn}{N}}$. The rank of a matrix $\bf A$ is denoted by $r({\bf A})$. The vector space spanned by vectors ${\bf x}_1, \ldots, {\bf x}_N,$ is denoted by $\mathcal{L}({\bf x}_1, \ldots, {\bf x}_N)$. The vector space spanned by the columns of $\bf A$ is denoted by $\mathcal{L}({\bf A})$. $\mathcal{L}({\bf A})\bot \mathcal{L}({\bf B})$ means that the column spaces of $\bf A$ and $\bf B$ are orthogonal to each other, i.e., ${\bf A}^H{\bf B}=0$. $\bf 0$ is an all-zero matrix or a vector whose dimension is understood from the context. $\delta(x)$ denotes the Kronecker-delta function, i.e., $\delta(0)=1$ and $\delta(x)=0$ for $x\neq0$.

\section{System Model}
\subsection{OFDM System Model}
Consider an OFDM system with $N$ subcarriers. Without loss of generality, we consider the zeroth OFDM symbol. The $N\negmedspace\times\negmedspace 1$ OFDM symbol vector at the transmitter, which consists of data and pilot symbols, is denoted by ${\bf x}$. The $l$th channel tap at the $n$th time instant is denoted by ${\tilde h}_l(n)$. The length of the cyclic prefix (CP) is $L$ samples (which is also the number of distinct resolvable multipaths of the DS channel).  The samples at the receiver, ${\tilde r(n)}$, are corrupted by additive white Gaussian noise (AWGN), ${\tilde w(n)}$. After discarding the CP, the $N\negmedspace\times\negmedspace 1$ received vector can be written as
\begin{equation}\label{matrixmpeqn}
{\bf \tilde r}  = \frac{1}{\sqrt{N}}{\bf \tilde H} {\bf F}^H {\bf x} + {\bf \tilde n}
\end{equation}
where 
$
{\bf \tilde r}= [{\tilde r(L)},\ldots,{\tilde r(L+N - 1)}]^T 
$, $
{\bf \tilde n} = [ {\tilde w}(L), \ldots, {\tilde w} (L+N-1) ]^T 
$ is the AWGN vector and 
$
{\bf \tilde H}(p,q) = {\tilde h}_{\left\langle p-q\right\rangle}(L+p)$ is the time-domain channel matrix \cite{pilot_JSP_07}.  An FFT is applied to ${\bf \tilde r}$ to yield 
\begin{equation}\label{t2f1}
\begin{array}{*{20}l}
   {\bf y}  = \frac{1}{{\sqrt N }}{\bf F\tilde r}   
    = \underbrace {\frac{1}{N}{\bf F\tilde H} {\bf F}^H }_{{\bf H} }{\bf x}  + \underbrace {\frac{1}{{\sqrt N }}{\bf F}{\bf \tilde n} }_{{\bf n} }  \\
\end{array}
\end{equation}
where $
{\bf y}  = [ {y (0)},\ldots, {y (N - 1)}]^T 
$ is the vector of demodulated output symbols. The input-output relation is therefore given by
${\bf y}  = {\bf H} {\bf x}  + {\bf n} \cdot$
\subsection{BEM Channel Model}
Let the $l$th-path channel vector be denoted as ${\bf {\tilde h}}_l  = [ {\tilde h_l \left( L \right), \ldots, \tilde h_l \left( {L + N - 1} \right)} ]^T 
$. This can be quite accurately characterized by a $Q$th order BEM as \cite{pilot_JSP_07}
 \begin{equation}\label{mpbem}{\bf \tilde h}_l \approx {\bf B} {\bf h}_l\end{equation} where 
 ${\bf  h}_l\negmedspace=\negmedspace[ h_l(0),\ldots, h_l(Q\!-\!1)]^T$ is the vector of BEM parameters and $\bf B$ is an $N\negmedspace\times\negmedspace Q$ BEM matrix. Note that the $N$ samples of the channel are characterized by $Q$ BEM parameters and $Q\ll N$.
 
\subsection {Pilot and Data Clusters}
The system has $N_P$ pilot clusters and $N_D$ data clusters, equispaced and interleaved with each other as shown in Fig. \ref{cluster}. The lengths of data and pilot clusters are $L_D = 2w_D+1$ and $L_P = 2w_P+1$, respectively. We denote the pilot and data vectors by the $N\negmedspace\times\negmedspace 1$ vectors $\bf p$ and $\bf d$, respectively. The quantity ${\bf p}(i)$ denotes the pilot symbol mapped to the $i$th subcarrier and is equal to 0 if the $i$th subcarrier is a data subcarrier. Likewise, ${\bf d}(i)$ denotes the data symbol mapped to the $i$th subcarrier and is equal to 0 if the $i$th subcarrier is a pilot subcarrier. Note that ${\bf x} = {\bf p}+{\bf d}$. Equation \eqref{t2f1} can be rewritten as 
\begin{equation}\label{pplusd}
{\bf y} = \frac{1}{N}{\bf F\tilde HF}^H {\bf p} + \frac{1}{N}{\bf F\tilde HF}^H {\bf d} + \frac{1}{{\sqrt N }}{\bf F \tilde n} \cdot 
\end{equation}
Let $P_b$ and $P_{\rm sep}$ denote the index of the middle subcarrier of the zeroth pilot cluster and  the distance between the middle subcarriers of two neighboring pilot clusters, respectively.  Since the pilot clusters are equispaced it follows that $N=N_PP_{\rm sep}$. Let ${\bf p}_1$ be an $N_PL_P\!\times\! 1$ pilot-index vector whose entries comprise  the indices of all pilot subcarriers in ascending order, i.e.,
\begin{equation}\label{p1}
   {{\bf p}_1  = } {\left[ {\left\langle {P_b  - w_P } \right\rangle ,} \right.} { \ldots ,} {P_b ,} \ldots,{\left\langle {P_b  + w_P } \right\rangle ,} { \ldots ,} {\left. {\left\langle {P_b \negthickspace +\negthickspace (N_P\negthickspace  -\negthickspace 1)P_{\rm sep}\negthickspace  +\negthickspace w_P } \right\rangle } \right]^T  } \cdot
\end{equation}
The pilot cluster consists of two parts. A centrally located observation cluster (OC) of length $O\!=\!2B_c\!+\!1$ is used for channel estimation. It is protected on either side by a guard band of length $G$ which prevents interference from the neighboring data clusters. Its length is dependent on the Doppler spread of the channel. The pilot cluster length is therefore given by $L_P\negmedspace=\negmedspace O\negmedspace+\negmedspace 2G$.
In all, $V\negmedspace=\negmedspace N_PO$ pilot subcarriers are used for channel estimation. The $V$ pilot subcarriers are accessed via $(2B_c\!+\!1)$ pilot-index vectors ${\bf p}_2^{(i)}, i\!=\!-B_c,\!\ldots\!, B_c$, where  ${\bf p}_2^{(i)}$ is the $i$th pilot-index vector. It comprises of the indices of the $i$th neighbour of the middle subcarrier in each pilot cluster and is defined as
\begin{equation}\label{p2i}
\begin{array}{*{20}r}
   {\bf p}_2^{\left( i \right)}\negthickspace\negthickspace &\negthickspace\negthickspace =\negthickspace\negthickspace&\negthickspace\negthickspace \left[ {\left\langle {P_b \negthickspace +\negthickspace i} \right\rangle ,\left\langle {P_b \negthickspace +\negthickspace i\negthickspace +\negthickspace P_{{\rm sep}} } \right\rangle \negthickspace, \ldots ,\negthickspace\left\langle {P_b\negthickspace  +\negthickspace i\negthickspace +\negthickspace \left( {N_P\negthickspace  -\negthickspace 1} \right)P_{{\rm sep}} } \right\rangle } \right]^T\negthickspace , \:\: &&{\left| i \right| \le B_c }  \cdot
\end{array}
\end{equation}
Concatenating all ${\bf p}_2^{(i)}$ into a master pilot-index vector ${\bf p}_2=\left[{\bf p}_2^{(-B_c)T}, \ldots, {\bf p}_2^{(B_c)T}\right]^T$, the observation pilot symbol vector used for channel estimation is the $V\negmedspace\times\negmedspace 1$ vector ${\bf \bar y}={\bf y}({\bf p}_2)$.
\subsection{BEM-Based OFDM System Model}
Consider the $P_b$th demodulated pilot symbol that belongs to the zeroth pilot cluster. Neglecting interference from data clusters and AWGN, it follows from \eqref{pplusd} that
\begin{equation}
{\bf y}\left( {P_b } \right) = \underbrace {\frac{1}{N}\sum\limits_{i =  - w_P }^{i = w_P } {{\bf p} \left( {P_b  + i} \right)\sum\limits_{l = 0}^{L - 1} {\sum\limits_{n = 0}^{N - 1} {\tilde h_l \left( {L + n} \right)e^{j2\pi \left( {P_b  + i} \right)\Delta f(n-l)} e^{ - j2\pi P_b \Delta fn} } } } }_{{\rm zeroth\: pilot\: cluster}} + \underbrace {\sum  \cdots  }_{{\rm first,} \ldots {\rm ,}\left( {N_P  - 1} \right){\rm \:st\: pilot\: cluster}}
\end{equation}
where $\Delta f\negthickspace=\negthickspace1/N$. Rearranging and interchanging the summations, we have
\begin{equation}\label{tmp}
{\bf y}\left( {P_b } \right) = \underbrace {\frac{1}{N}\sum\limits_{l = 0}^{L - 1} {\sum\limits_{i =  - w_P }^{w_P } {{\bf p} \left( {P_b  + i} \right)e^{ - j2\pi \left( {P_b  + i} \right)\Delta fl} \sum\limits_{n = 0}^{N - 1} {\tilde h_l \left( {L + n} \right)e^{j2\pi i\Delta fn} } } } }_{{\rm zeroth\: pilot\: cluster}} + \underbrace {\sum  \cdots  }_{{\rm first,} \ldots {\rm ,}\left( {N_P  - 1} \right){\rm st\: pilot\: cluster}}\cdot
\end{equation}
Define the inverse fast Fourier transform (IFFT) matrix of ${\tilde{\bf h}}_l$ as
\begin{equation}
{\bf H}_F^{(l)}=(1/N){\bf F}^H{\tilde{\bf h}}_l\cdot
\end{equation}
The above equation represents the Doppler spectrum of the $l$th channel tap. Equation \eqref{tmp} can then be rewritten in a concise form as
\begin{equation}\label{ypb}
{\bf y}\left( {P_b } \right)\negthickspace =\negthickspace\negthickspace \sum\limits_{l = 0}^{L - 1}\negthickspace\negthickspace \begin{array}{*{20}l}
{\bf p} \left( {P_b \negthickspace -\negthickspace w_P } \right)e^{ - j2\pi \left( {P_b  - w_P } \right)\Delta fl} {\bf H}_F^{\left( l \right)}\negthickspace \left( {N\negthickspace -\negthickspace w_P } \right) +  \cdots  + {\bf p} \left( {P_b  + w_P } \right)e^{ - j2\pi \left( {P_b \negmedspace +\negmedspace w_P } \right)\Delta fl} {\bf H}_F^{\left( l \right)} \left( {w_P } \right)\\
+{\bf p} \left( {P_b \negthickspace+\negthickspace P_{\rm sep}\negthickspace -\negthickspace w_P } \right)e^{ - j2\pi \left( {P_b  +P_{\rm sep}- w_P } \right)\Delta fl} {\bf H}_F^{\left( l \right)}\negthickspace \left( {P_{\rm sep}\negthickspace -\negthickspace w_P } \right) +  \cdots  \\
\end{array} \cdot
\end{equation}
Similarly, the $(P_b+P_{\rm sep})$st demodulated pilot subcarrier of the first pilot cluster, can be written as
\begin{equation}\label{ypbpsep}
{\bf y}\left( {P_b \negmedspace +\negmedspace P_{{\rm sep}} } \right)\negmedspace =\negmedspace
\sum\limits_{l = 0}^{L - 1} \begin{array}{*{20}l} 
{\bf p} \left( {P_b\negmedspace  +\negmedspace P_{{\rm sep}}\negmedspace  -\negmedspace w_P } \right)e^{ - j2\pi \left( {P_b\negmedspace  + \negmedspace P_{{\rm sep}}\negmedspace  -\negmedspace w_P } \right)\Delta fl} {\bf H}_F^{\left( l \right)} \left( {N - w_P } \right) +  \cdots  \\
+ {\bf p} \left( {P_b \negmedspace +\negmedspace P_{{\rm sep}}\negmedspace+\negmedspace w_P } \right)e^{ - j2\pi \left( {P_b \negmedspace + \negmedspace P_{{\rm sep}} \negmedspace +\negmedspace w_P } \right)\Delta fl} {\bf H}_F^{\left( l \right)} \left( {w_P } \right)\\ 
+  {\bf p} \left( {P_b \negmedspace +\negmedspace 2P_{{\rm sep}}\negmedspace-\negmedspace w_P } \right)e^{ - j2\pi \left( {P_b \negmedspace + \negmedspace 2P_{{\rm sep}} \negmedspace -\negmedspace w_P } \right)\Delta fl} {\bf H}_F^{\left( l \right)} \left( P_{\rm sep}-{w_P } \right) + \ldots  \cdot
\end{array}
\end{equation}

Before we proceed further, we present a physical interpretation of \eqref{ypb} and \eqref{ypbpsep}. The quantity \newline${\bf p} \left( {P_b \negthickspace -\negthickspace w_P } \right)e^{ - j2\pi \left( {P_b  - w_P } \right)\Delta fl} {\bf H}_F^{\left( l \right)}\negthickspace \left( {N\negthickspace -\negthickspace w_P } \right)$ represents the response of the $l$th path of the channel at Subcarrier $P_b$ due to the excitation by Subcarrier $(P_b-w_P)$. The quantity $e^{ - j2\pi \left( {P_b  - w_P } \right)\Delta fl}$ is due to the delay of the $l$th path channel. The quantity ${\bf H}_F^{\left( l \right)}\negthickspace \left( {N\negthickspace -\negthickspace w_P } \right)$ represents the spillage/leakage of the $l$th path Doppler spectrum. It can also be viewed as the intercarrier interference (ICI) contribution of Subcarrier $(P_b-w_P)$ on Subcarrier $P_b$ via the $l$th channel. It therefore follows that \eqref{ypb} and \eqref{ypbpsep} can be viewed as the {\it sum of the frequency-domain convolutions of the Doppler spread of each time-shifted channel path with the pilot cluster}. A detailed mathematical discussion and pictorial depiction of this interpretation was given by Scneiter in \cite{phillip}. 

Observe that  ${\bf H}_F^{\left( l \right)} \left( {N - w_P } \right)$ is associated with ${\bf p} \left( {P_b  - w_P } \right)$ in \eqref{ypb}, while in \eqref{ypbpsep} it is associated with ${\bf p} \left( {P_b  \negmedspace +\negmedspace P_{\rm sep}\negmedspace-\negmedspace w_P } \right)$. Let the  row-vector ${\bf p}_c^{(i)}=\left[{\bf p}(P_b+iP_{\rm sep}-w_P),\ldots,{\bf p}(P_b+iP_{\rm sep}+w_P)\right]$ denote the $i$th pilot cluster. The pilot-pattern matrix, ${\bf P}_{{\rm pat}}$, is obtained by stacking all the pilot-cluster row vectors and $\bf Q$ is a pilot-indices matrix whose $i$th row comprises of the indices of the $i$th pilot cluster, i.e.,
\begin{equation}\label{pandq}
\begin{array}{*{20}r}
{\bf P}_{{\rm pat}}  = \left[ {\begin{array}{*{20}c}
   {{\bf p}_c^{\left( 0 \right)} }  \\
    \vdots   \\
   {{\bf p}_c^{\left( {N_P  - 1} \right)} }  \\
\end{array}} \right], & {\bf Q} = \left[ {\begin{array}{*{20}l}
   {P_b  - w_P ,} &  \ldots  & {P_b  + w_P }  \\
   {} &  \vdots  & {}  \\
   {P_b  + \left( {N_P  - 1} \right)P_{{\rm sep}}  - w_P ,} &  \ldots  & {P_b  + \left( {N_P  - 1} \right)P_{{\rm sep}}  + w_P }  \\
\end{array}} \right] \cdot
\end{array}
\end{equation}
\begin{figure}[t]
     \centering 
      \vspace {0.0in} 
        \includegraphics [width = 8cm,height = 3cm] {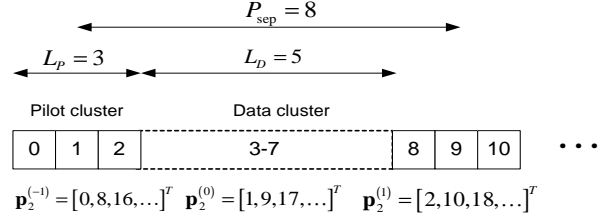}
         \caption{An example of pilot and data clusters.}\label{cluster}
         \vspace{0.0in} 
\end{figure}

For the example in Fig. \ref{cluster}, the matrices ${\bf P}_{\rm pat}$ and $\bf Q$ are given as
\[
\begin{array}{*{20}c}
   {{\bf P}_{{\rm pat}}  = \left[ {\begin{array}{*{20}c}
   {\begin{array}{*{20}c}
   {{\bf p}\left( 0 \right)} & {{\bf p}\left( 1 \right)} & {{\bf p}\left( 2 \right)}  \\
\end{array}}  \\
   {\begin{array}{*{20}c}
   {{\bf p}\left( 8 \right)} & {{\bf p}\left( 9 \right)} & {{\bf p}\left( {10} \right)}  \\
\end{array}}  \\
    \vdots   \\
\end{array}} \right]}, & {{\bf Q} = \left[ {\begin{array}{*{20}c}
   {\begin{array}{*{20}c}
   0 & 1 & 2  \\
\end{array}}  \\
   {\begin{array}{*{20}c}
   8 & 9 & {10}  \\
\end{array}}  \\
    \vdots   \\
\end{array}} \right]}  \\
\end{array}\cdot
\]

We are now in a position to present the expression for ${\bf y}({\bf p}_2^{(0)})$. To do so, we define the following
\begin{equation}\label{manyeq}
\begin{array}{*{20}l}
{\underline{\bm\theta}}_l  &=& \left[ {\begin{array}{*{20}c}
   {e^{ - j2\pi {\bf Q}\Delta fl}  \odot {\bf P}_{{\rm pat}} }  \\
   {e^{ - j2\pi {\bf Q}\Delta fl}  \odot {\bf P}_{{\rm pat}} }  \\
\end{array}} \right]\\
{\bf R}_l^{\left( i \right)}  &=& {\underline{\bm\theta}} _l \left( {i:N_P  + i - 1,:} \right)\\
{{\bm\theta}} _l  &=& \left[ {\begin{array}{*{20}c}
   {{\bf R}_l^{\left( 0 \right)} ,} & {{\bf R}_l^{\left( 1 \right)} ,} & { \ldots ,} & {{\bf R}_l^{\left( {N_P  - 1} \right)} }  \\
\end{array}} \right]\\
   {\bf W}^{(i)} &=& \frac{1}{N}\left[ {\begin{array}{*{20}c}
   {{\bf F}^H \left( {\left\langle {N + ({\bf p}_1(0)-P_b-i) } \right\rangle,: } \right)}  \\
   {{\bf F}^H \left( {\left\langle {N + ({\bf p}_1(1)-P_b-i) } \right\rangle,: } \right)}  \\
    \vdots   \\
   {{\bf F}^H \left( {\left\langle {N + ({\bf p}_1(N_PL_P-1)-P_b-i)} \right\rangle,: } \right)}  \\
\end{array}} \right]  \\
{\bf h} &=& \left[ {{\bf h}_0^T,  \ldots, {\bf h}_{L - 1}^T } \right]^T \cdot\\
\end{array}
\end{equation}

Observe that the rows of the matrices ${{\bf R}_l^{\left( 0 \right)} ,\:} {{\bf R}_l^{\left( 1 \right)} ,\:}  { \ldots ,\:} {{\bf R}_l^{\left( {N_P  - 1} \right)} },$ are circularly shifted with respect to one another, i.e., for instance we have 
\begin{equation}\label{circ}{\bf R}_l^{\left( 0 \right)}(1,:)={\bf R}_l^{\left( 1 \right)}(0,:)\cdot
\end{equation} Furthermore, we observe that
\begin{equation}\label{ex1}
\left[   {\bf H}_F^{\left( l \right)} \left( N-w_P \right), \ldots, {\bf H}_F^{\left( l \right)} \left( w_P \right), {\bf H}_F^{\left( l \right)} \left( P_{\rm sep}-w_P \right), \ldots \right]^T  = {\bf W}^{\left( 0 \right)} {\bf \tilde h}_l  \simeq {\bf W}^{\left( 0 \right)} {\bf Bh}_l
\end{equation} 
\begin{equation}\label{ex2}
{\bm\theta} _l  = \left[ {\begin{array}{*{20}l}
   {{\bf p}\left( {P_b  - w_P } \right)e^{ - j2\pi \left( {P_b  - w_P } \right)\Delta fl} } &  \ldots   & {{\bf p}\left( {P_b  + P_{{\rm sep}}  - w_P } \right)e^{ - j2\pi \left( {P_b  + P_{{\rm sep}}  - w_P } \right)\Delta fl} } &  \cdots   \\
   {{\bf p}\left( {P_b  + P_{{\rm sep}}  - w_P } \right)e^{ - j2\pi \left( {P_b  + P_{{\rm sep}}  - w_P } \right)\Delta fl} } &  \ldots  & {{\bf p}\left( {P_b  + 2P_{{\rm sep}}  - w_P } \right)e^{ - j2\pi \left( {P_b  + 2P_{{\rm sep}}  - w_P } \right)\Delta fl} } &  \ldots   \\
    \vdots  & {} &  \vdots   & {}   \\
\end{array}} \right]\cdot
\end{equation}
As already discussed,  \eqref{ex1} represents the Doppler spectrum of the (time-aligned) $l$th channel tap. 
Recall the fact that interference from data clusters and noise is neglected for the moment and ${\bf y}({\bf p}_2^{(0)})=\left[{\bf y}(P_b), {\bf y}(P_b+P_{\rm sep}), \ldots, \right]^T$. It follows from \eqref{ypb}, \eqref{ypbpsep}, \eqref{manyeq}, \eqref{ex1} and \eqref{ex2} that
\begin{equation}\label{bemsysmodel}
{\bf y}({\bf p}_2^{(0)})={\bf H}\left( {{\bf p}_2^{(0)} ,:} \right){\bf p} = \left[ {\begin{array}{*{20}c}
   {{\bm\theta} _0 {\bf W}^{\left( 0 \right)} {\bf B},} & { \cdots ,} & {{\bm\theta} _{L - 1} }  \\
\end{array}{\bf W}^{\left( 0 \right)} {\bf B}} \right]{\bf h}\cdot
\end{equation}
Considering the interference due to data clusters and AWGN, recalling that ${\bar{\bf y}}={\bf y}({\bf p}_2)$, we have
\begin{equation}\label{yeqhp}
{\bf \bar y} = {\boldsymbol{\mathcal P}}{\bf h}+{\bf n}_1
\end{equation}
where ${\bf n}_1$ is the interference due to data clusters and noise and 
\begin{equation}\label{P}
{\boldsymbol{\mathcal P}} = \left[ {\begin{array}{*{20}c}
   {\underbrace {\begin{array}{*{20}c}
   {{\bm \theta} _0 {\bf W}^{\left( { - B_c } \right)} {\bf B}}  \\
    \vdots   \\
   {{\bm \theta} _0 {\bf W}^{\left( {B_c } \right)} {\bf B}}  \\
\end{array}}_{{\bm \phi} _0 }} & {\begin{array}{*{20}c}
    \ldots   \\
    \vdots   \\
    \ldots   \\
\end{array}} & {\underbrace {\begin{array}{*{20}c}
   {{\bm \theta} _{L - 1} {\bf W}^{\left( { - B_c } \right)} {\bf B}}  \\
    \vdots   \\
   {{\bm \theta} _{L - 1} {\bf W}^{\left( {B_c } \right)} {\bf B}}  \\
\end{array}}_{{\bm \phi} _{L - 1} }}  \\
\end{array}} \right]
\end{equation}
is the channel estimation matrix of dimension $V\negmedspace\times\negmedspace LQ$. In the above equation,
${\bm \phi}_l$ is associated with the $l$th multipath and has the form
\begin{equation}\label{phil}
{\bm \phi} _l  = \underbrace{\left[ {\begin{array}{*{20}c}
   {{\bm \theta} _l } & {} & {}  \\
   {} &  \ddots  & {}  \\
   {} & {} & {{\bm \theta} _l }  \\
\end{array}} \right]}_{{\bar {\bm \theta}}_l}\underbrace {\left[ {\begin{array}{*{20}c}
   {{\bf W}^{\left( { - B_c } \right)} {\bf B}}  \\
    \vdots   \\
   {{\bf W}^{\left( {B_c } \right)} {\bf B}}  \\
\end{array}} \right]}_{\bf E} \cdot
\end{equation}
The purpose of this paper is to determine the conditions on the BEM matrix $\bf B$ and pilot clusters ${\bf p}_c^{(.)}$ such that $\boldsymbol{\mathcal P}$ is of full column rank. 

The matrix $\boldsymbol {\mathcal P}$ defined in \eqref{P} and the one defined in \cite[(14)]{pilot_JSP_07} are permuted versions of each other and have the same rank. The interchange of rows (columns) of a matrix does not change its rank. Equation \eqref{yeqhp} is a permuted version of (14) in \cite{pilot_JSP_07}. More specifically, $\bar {\bf y}$ in this paper is defined as
\begin{equation}\label{ybarthispaper}
\begin{array}{*{20}l}
   {\bar{\bf y}  = } \left[ {\bf y}(P_b \! -\! B_c) , {\ldots }, {\bf y}(P_b\!+\!(N_P\!-\!1)P_{\rm sep}\!  -\! B_c),
   \ldots, {\bf y}(P_b) , {\ldots }, {\bf y}(P_b \! +\! B_c), \ldots , {\bf y}(P_b\!+\!(N_P\!-\!1)P_{\rm sep}\!  +\! B_c) \right]^T 
\end{array}\end{equation}
while ${\bf y}^{(p)}$ in \cite[(14)]{pilot_JSP_07} is defined as 
\begin{equation}\label{yofp}
\begin{array}{*{20}l}
   {{\bf y}^{(p)}  = } \left[ {\bf y}(P_b \! -\! B_c) , {\ldots }, {\bf y}(P_b \! +\! B_c),
   \ldots,{\bf y}(P_b\!+\!(N_P\!-\!1)P_{\rm sep}\!  -\! B_c),
   \ldots , {\bf y}(P_b\!+\!(N_P\!-\!1)P_{\rm sep}\!  +\! B_c) \right]^T  \cdot\\
\end{array}\end{equation}
Observe that \eqref{ybarthispaper} and \eqref{yofp} are permuted versions of each other. Similarly, ${\bf h}$ in this paper and $\bf h$ in \cite[(4)]{pilot_JSP_07} are also permuted versions of each other.
\section{Full Column-Rank Condition of ${\boldsymbol{\mathcal P}}$}
Recall that ${\boldsymbol{\mathcal P}}$ is a $V\times QL$ matrix and the full column rank necessitates that $r({\boldsymbol{\mathcal P}})=QL$. The matrices ${\bm\phi}_i,0\leq i\leq L-1,$ are of dimension $V\times Q$. It therefore follows from \eqref{P} that 
\begin{equation}\label{conditions}
\begin{array}{*{20}l}
r\left({\bm \phi}_i\right) &=& Q,&0\leq i\leq L-1,\\{\mathcal{L}}\left({\bm \phi}_j\right) &\bot& {\mathcal{L}}\left({\bm \phi}_i\right),& i\neq j 
\end{array}
\end{equation}
needs to be satisfied for ${\boldsymbol{\mathcal P}}$ to be of full column rank. To satisfy the second condition in the above equation, it follows from \eqref{phil} that the relationship
\begin{equation}\label{tetalhtetal}
{\bm\theta}_i^H{\bm\theta}_j = 0, \:\:\:\: i\neq j
\end{equation} needs to be satisfied. We now present the condition that needs to be satisfied by a BEM to ensure that ${\boldsymbol{\mathcal P}}$ is of full column rank. The dimensions of the matrices ${{\bar {\bm \theta}}_l}$ and $\bf E$ are $V\times VL_P$ and $VL_P\times Q$, respectively. Generally, we have the relationship $V\gg Q$. Recall that ${\bm \phi}_l={\bar \theta}_l{\bf E}$. To ensure that the first condition in \eqref{conditions} is satisfied, it therefore becomes necessary that a BEM satisfies
\begin{equation}\label{ranke}
r\left({\bf E}\right) = Q
\end{equation}
which will be termed as the BEM condition (BEMC). The BEMC necessary condition for the matrix ${\boldsymbol{\mathcal P}}$ to be full column rank means that the BEM matrix $\bf B$ containing the independent $Q$ basis functions should note lose rank when it is projected in the Doppler (frequency) domain, where the observations are collected.

Before we proceed further, we present the rank-nullity theorem. Consider a matrix $\bf B$ such that $r\left({\bf B}\right)=Q$ and ${\bf b}_0, \ldots, {\bf b}_{Q-1},$ be the basis vectors that span the column space of $\bf B$. If none of ${\bf b}_0, \ldots, {\bf b}_{Q-1},$ lie in the null space of another matrix $\bf A$, then the rank-nullity theorem \cite{rnt} states that $r\left({\bf AB}\right)=Q$ and ${\bf Ab}_0, \ldots, {\bf Ab}_{Q-1},$ span the column space (basis vectors) of ${\bf AB}$.
%
\section{Pilot-Pattern Design}
The design of pilot patterns should be such that \eqref{conditions} and \eqref{tetalhtetal} are satisfied. In this section, we outline {\it one possible method of pilot-pattern design, that satisfies the conditions in \eqref{conditions} and \eqref{tetalhtetal}, thus ensuring the full column-rank condition of the channel estimation matrix}. The design example is presented for both DS-OFDM and DS-MIMO-OFDM systems.
\subsection{DS-OFDM Systems}
We now design the pilot pattern matrix, ${\bf P}_{\rm pat}$, such that $\boldsymbol {\mathcal P}$ is of full column rank for a DS-OFDM system. Initially, we concentrate on the design of pilot patterns such that the second condition in \eqref{conditions} is satisfied. Towards the end of the section, we concentrate on the design of pilot patterns such that the first condition in \eqref{conditions} is also satisfied. We begin by examining ${\bf R}_l^{(0)}$. From \eqref{pandq}, it follows that the $r$th column of $\bf Q$ can be written as
\begin{equation}\label{Q}
{\bf Q}(:,r) = {\bf Q}(:,0)+r{\bf 1}
\end{equation}
where $\bf 1$ is a column vector with all entries equal to Unity. It therefore follows that $r({\bf Q})=2$. Recall from \eqref{manyeq} that ${\bf R}_l^{(0)}={\underline{\bm{\theta}}}_l \left( {0:N_P  - 1,:} \right) = e^{ - j2\pi {\bf Q}\Delta fl}\odot {\bf P}_{{\rm pat}}$. Now consider what happens when all entries of ${\bf P}_{\rm pat}$ are equal to Unity. It follows from \eqref{manyeq} and \eqref{Q} that all columns of ${\underline{\bm{\theta}}}_l(0:N_P-1,:)$ in this case are scaled versions of each other, i.e.,
\begin{equation}\label{tmp1}
\begin{array}{*{20}c}
   {\underline{\bm{\theta}}} _l \left( {0:N_P  - 1,:} \right) &=& \left[ {\begin{array}{*{20}c}
   {e^{ - j2\pi {\bf Q}\left( {:,0} \right)\Delta fl} } &  \ldots  & {e^{ - j2\pi r\Delta fl} e^{ - j2\pi {\bf Q}\left( {:,0} \right)\Delta fl} } &  \ldots   \\
\end{array}} \right]  \\
\end{array}
\end{equation}

Define an $N_P\times 1$ vector as ${\bf f}^{(i)}=\left[1, e^{-j2\pi i \Delta f_1}, \ldots, e^{-j2\pi i \Delta f_1(N_P-1)}\right]^T$, where $\Delta f_1 = (1/N_P)$. The quantity ${\bf f}^{(i)}$ can be considered as the $i$th harmonic of a $N_P\times 1$ vector whose fundamental discrete frequency is $\Delta f_1$. We observe that ${\bf f}^{(i)H}{\bf f}^{(j)}=0, i\neq j, 0\leq i,j \leq N_P-1$. Recall that the rows of $\bf Q$ differ by $P_{\rm sep}$, $N=N_PP_{\rm sep}$ and $\Delta f=1/N$. From the definition of $\bf Q$ in \eqref{pandq}, it follows that
\begin{equation}\label{spanofl}
\begin{array}{*{20}c}
   e^{ - j2\pi {\bf Q}\left( {:,0} \right)\Delta fl}  &=& e^{ - j2\pi \left( {P_b  - w_P } \right)\Delta fl} {\bf f}^{\left( l \right)}   \\
   {\mathcal L}\left( {e^{ - j2\pi {\bf Q}\left( {:,0} \right)\Delta fl} } \right) &=& {\mathcal L}\left( {{\bf f}^{\left( l \right)} } \right)  \cdot\\
\end{array}
\end{equation}
Under the assumption that all pilot symbols are equal to 1 (entries of ${\bf P}_{\rm pat}$ is equal to Unity), it follows from \eqref{tmp1} and \eqref{spanofl} that
\[
\begin{array}{*{20}c}
   {r\left( {{\bf R}_l^{\left( 0 \right)} } \right) = 1,} & {{\mathcal L}\left( {{\bf R}_l^{\left( 0 \right)} } \right) =   {\mathcal L}\left( {\bf f}^{(l)} \right)}  \\
\end{array} \cdot
\]
Furthermore, recall from \eqref{circ} that the rows of the various matrices, ${\bf R}_l^{(i)}, 0\leq i\leq N_P-1,$ are all circularly shifted with respect to one another. It therefore follows that
\[
\begin{array}{*{20}c}
   {r\left( {{\bf R}_l^{\left( i \right)} } \right) = 1,} & {{\mathcal L}\left( {{\bf R}_l^{\left( 0 \right)} } \right) =  \cdots  = {\mathcal L}\left( {{\bf R}_l^{\left( {N_P  - 1} \right)} } \right) = {\mathcal L}\left( {\bf f}^{(l)} \right)}  \\
\end{array} \cdot
\]

Next, we design ${\bf P}_{\rm pat}$ such that the following conditions are satisfied
\begin{equation}\label{2c}
\begin{array}{*{20}c}
   {{\mathcal L}\left( {{\bf R}_l^{\left( 0 \right)} } \right) =  \ldots  = {\mathcal L}\left( {{\bf R}_l^{\left( {N_P  - 1} \right)} } \right),} & {0 \le l \le L - 1}  \\
   {{\mathcal L}\left( {{\bf R}_l^{\left( 0 \right)} } \right) \bot  \ldots  \bot {\mathcal L}\left( {{\bf R}_{L - 1}^{\left( 0 \right)} } \right)} & {}  \\
\end{array}
\end{equation}
which would ensure that \eqref{tetalhtetal} will be satisfied. This ensures that the second condition in \eqref{conditions} is also satisfied. The matrix ${\underline{\bm\theta}_l} $ is the concatenation of the matrix, $e^{ - j2\pi {\bf Q}\Delta fl}  \odot {\bf P}_{{\rm pat}} $, with itself, one on top of another. From \eqref{pandq}, \eqref{manyeq} and \eqref{spanofl}, it follows that ${\underline{\bm\theta}_l}(0:N_P-1,:) $ can be written as
\begin{equation}\label{underlinetetal}
{\bf R}_l^{(0)}={\underline{\bm\theta}_l(0:N_P-1,:)} = \left[ {\begin{array}{*{20}c}
   {e^{ - j2\pi {\bf Q}\left( {:,0} \right)\Delta fl} {\bf P}_{{\rm pat}} \left( {:,0} \right)} &  \ldots  & {e^{ - j2\pi {\bf Q}\left( {:,L_P  - 1} \right)\Delta fl} {\bf P}_{{\rm pat}} \left( {:,L_P  - 1} \right)}  \\
\end{array}} \right]\cdot
\end{equation}

\begin{figure}[t]
     \centering 
      \vspace {0.0in} 
        \includegraphics [width = 12cm,height = 8cm] {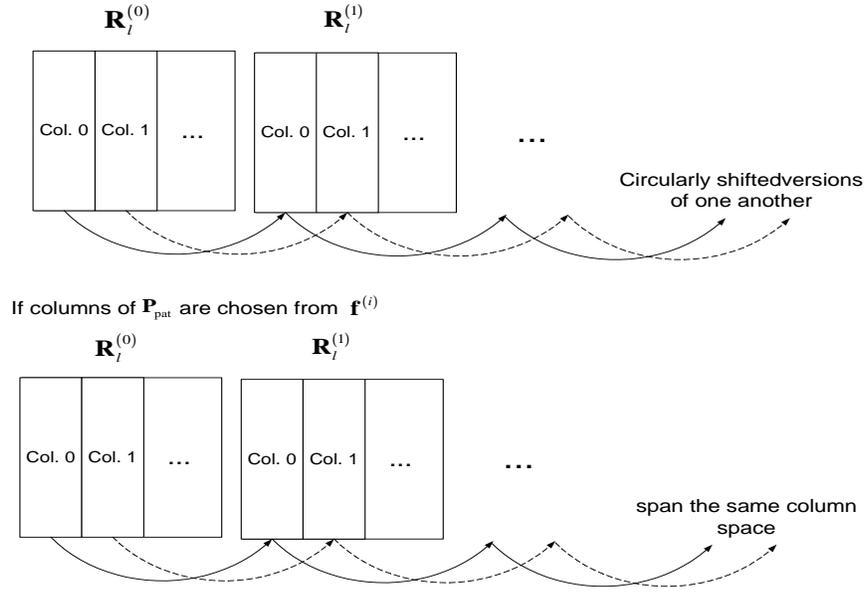}
         \caption{Properties of Matrices ${\bf R}_l^{(i)}, 0\leq i\leq N_P-1$.}\label{matrix1}
         \vspace{0.0in} 
\end{figure}

At this juncture, we recall the following:
\begin{itemize}
	\item It follows from \eqref{manyeq} that the rows of the various matrices ${\bf R}_l^{(i)}, 0\leq i \leq N_P-1,$ are all circularly shifted with respect to one another, as was depicted in \eqref{circ}. This is depicted in the upper part of Fig. \ref{matrix1}.
	\item If the columns of ${\bf P}_{\rm pat}$ are chosen from ${\bf f}^{(i)}, 0\leq i\leq N_P-1,$ then the first condition in \eqref{2c} is easily satisfied. We explain this by way of an example. Let ${\bf P}_{\rm pat}(:,0)={\bf f}^{(1)}$. It follows from \eqref{manyeq}, \eqref{spanofl} that ${\bf R}_l^{(0)}(:,0)={\underline{\bm \theta}}_l(0:N_P-1,0)= e^{ - j2\pi \left( {P_b  - w_P } \right)\Delta fl} {\bf f}^{\left( l+1 \right)}$. Similarly, ${\bf R}_l^{(1)}(:,0)={\underline{\bm \theta}}_l(1:N_P,0)= e^{-j2\pi\Delta f_1}e^{ - j2\pi \left( {P_b  - w_P + P_{\rm sep}} \right)\Delta fl} {\bf f}^{\left( l+1 \right)}$. Hence, it follows that ${\mathcal L}\left({\bf R}_l^{(0)}(:,0)\right)={\mathcal L}\left({\bf R}_l^{(1)}(:,0)\right)={\mathcal L}\left({\bf f}^{(l+1)}\right)$. This holds good for other columns of ${\bf R}_l^{(i)}, 0\leq i\leq N_P-1,$ as well. This is depicted in the lower part of Fig. \ref{matrix1}.
\end{itemize}

\begin{figure}[t]
     \centering 
      \vspace {0.0in} 
        \includegraphics [width = 12cm,height = 8cm] {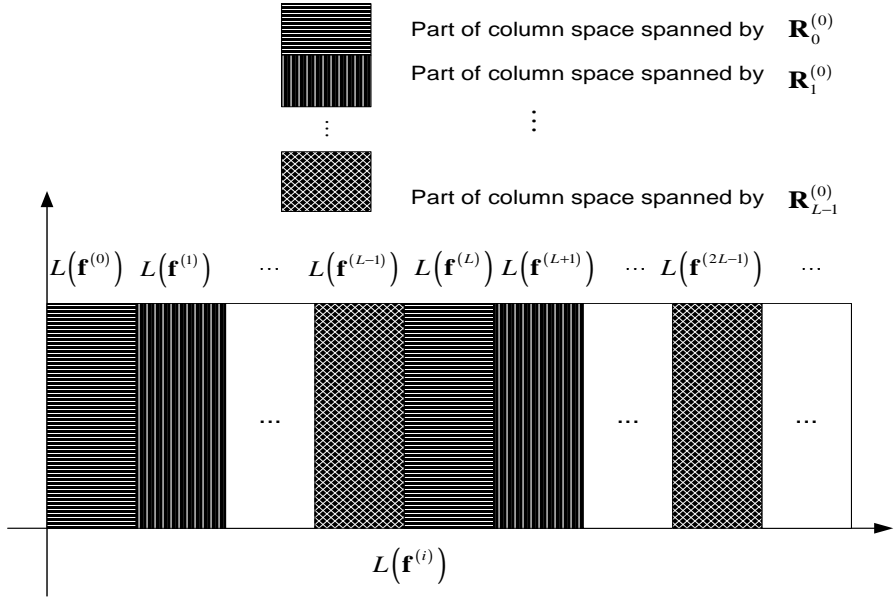}
         \caption{Column spaces of Matrices ${\bf R}_i^{(0)}, 0\leq i\leq L-1,$ are interleaved with one another.}\label{matrix2}
         \vspace{0.0in} 
\end{figure}

In order to satisfy the second condition in \eqref{2c}, we need to distribute the vectors ${\bf f}^{(i)}$ among the columns of ${\bf P}_{\rm pat}$. What should be the mapping strategy? Define ${\bar {\mathcal L}}(a,b)={\mathcal L}({\bf f}^{(a)}, {\bf f}^{(b)})$ and let ${\bf P}_{\rm pat}(:,r)={\bf f}^{(g_r)}$. We observe that ${\bar {\mathcal L}}(a) \bot {\bar {\mathcal L}}(a), a\neq b$. It follows from \eqref{underlinetetal} that
\begin{equation}\label{manyls}
\begin{array}{*{20}c}
   {\bf R}_0^{\left( 0 \right)}  &=& {\bar{\mathcal L}}\left( {g_{0, \ldots ,} g_{L_P  - 1} } \right)  \\
    &\vdots&   \\
   {\bf R}_l^{\left( 0 \right)}  &=& {\bar{\mathcal L}}\left( l+{g_{0, \ldots ,} l+g_{ L_P  - 1} } \right)  \\
    &\vdots&   \\
   {\bf R}_{L - 1}^{\left( 0 \right)}  &=& {\bar{\mathcal L}}\left( L-1+{g_{0, \ldots ,}L-1+ g_{L_P  - 1} } \right)  \\
\end{array}
\end{equation}
The above equation essentially says that the column space spanned by the $r$th column of ${\bf R}_l^{(0)}$ is
\[
{\mathcal L}\left({\bf R}_l^{(0)}\right) = {\bar{\mathcal L}}\left(l+g_r\right)\cdot
\]
In all, the $L$ matrices ${\bf R}_l^{(0)}, 0\leq l\leq L-1,$ have $L_PL$ columns. To satisfy the second condition in \eqref{2c}, it becomes necessary to select $g_r, 0\leq r \leq L_P-1,$ (select columns of ${\bf P}_{\rm pat}$) such that all the $L_PL$ columns of ${\bf R}_l^{(0)}, 0\leq l\leq L-1,$ are orthogonal to each other. In other words,
\begin{equation}
{\bar{\mathcal L}} \left(l_1+g_{r1}\right) \bot {\bar{\mathcal L}} \left(l_2+g_{r2}\right), \:\:\:\: 0\leq l_1,l_2 \leq L-1, \:\:0\leq {r_1},r_2 \leq L_P-1,\:\:\left(l_1, r_1\right)\neq \left(l_2, r_2\right)\cdot
\end{equation}
The above condition can easily be satisfied by the mapping
\begin{equation}
g_r = rL
\end{equation}
which implies that designing ${\bf P}_{\rm pat}$ as
\begin{equation}\label{ppatdsofdm}
\boxed{{\bf P}_{\rm pat} = \left[{\bf f}^{(0)},{\bf f}^{(L)}, \ldots, {\bf f}^{((L_p-1)L)}\right]}
\end{equation}
ensures that
\begin{equation}\label{lthetalprev}
\begin{array}{*{20}c}
   {{\mathcal L}\left( {{\bm\theta} _l } \right) = {\mathcal L}\left( {{\bf R}_l^{\left( 0 \right)} } \right) =  \cdots  = {\mathcal L}\left( {{\bf R}_l^{\left( {N_P  - 1} \right)} } \right) = {\mathcal L}\left( {{\bf f}^{\left( l \right)} ,{\bf f}^{\left( {{ L} + l} \right)} , \cdots ,{\bf f}^{\left( {\left( {L_P  - 1} \right)L+l} \right)} } \right)}  \\
   {\bm\theta}_i^H{\bm\theta}_j = 0, \:\:\:\: i\neq j  \cdot
\end{array}
\end{equation}
Consequently, the design as per \eqref{ppatdsofdm} ensures the full column-rank condition of $\boldsymbol{\mathcal P}$. From the above equation, it follows that the column spaces of Matrices ${\bf R}_l^{(0)}, 0\leq l \leq L-1,$ are {\it interleaved} with one another, as is depicted in Fig. \ref{matrix2}.

We now concentrate on the first condition in \eqref{conditions}, i.e., $r\left({\bm \phi}_i\right) = Q,0\leq i\leq L-1$. Matrix $\bf E$ is associated with the choice of BEM. The quantities ${{\bar {\bm \theta}}_l},0\leq l\leq L-1,$ are associated with the pilot-pattern design and ${\bm \phi}_l={\bar{\bm{\theta}}_l}{\bf E}$. It will be shown in the appendix and in Sec. VI, for the BEM models considered in this paper and the design of pilot pattern as per \eqref{ppatdsofdm}, none of the basis vectors of $\bf E$ lie in the null space of ${{\bar {\bm \theta}}_l},0\leq l\leq L-1$, i.e.,
\begin{equation}
{{\bar {\bm \theta}}_l}{\bf e}_Q \neq {\bf 0}, q=0,\ldots,Q-1,\:\:\:\:l=0,\ldots,L-1 \cdot
\end{equation}
This condition will be called as rank-nullity BEM condition (RNC-BEM).
This implies that if  the BEMC and RNC-BEM are satisfied, the rank-nullity theorem ensures that the first condition in \eqref{conditions}, i.e., $r\left({\bm \phi}_i\right) = Q,0\leq i\leq L-1,$ is also satisfied. It therefore follows that the pilot-pattern matrix designed as per \eqref{ppatdsofdm} ensures $\boldsymbol {\mathcal P}$ is of full column rank. In the appendix, we derive the RNC-BEM conditions in more detail for DS-OFDM systems. The derivations for DS-MIMO-OFDM systems can be pursued in a similar manner and is therefore not dealt in this paper.
 
Note that the $N_P\times 1$ vector ${\bf f}^{(i)}$ is the $i$th harmonic of a complex exponential vector with fundamental frequency $\Delta f_1=(1/N_P)$. The various ${\bf f}^{(.)}$ constitute the columns of ${\bf P}_{\rm pat}$ and these ensure that  ${{\mathcal L}\left( {{\bm\theta} _l } \right)={\mathcal L}\left( {{\bf f}^{\left( l \right)} ,{\bf f}^{\left( {{ L} + l} \right)} , \cdots ,{\bf f}^{\left( {\left( {L_P  - 1} \right)L+l} \right)} } \right)}$. In all, $L_PL$ vectors ${\bf f}^{(i)}, i=0, \ldots, L_PL-1,$ are associated with the $L$ matrices ${\bm \theta}_l,l=0,\ldots,L-1$. In order to ensure that ${\bm\theta}_i^H{\bm\theta}_j = 0, i\neq j,$ is satisfied, it becomes necessary that ${\bf f}^{(i)H}{\bf f}^{(j)}=0, i\neq j, 0\leq i,j \leq L_PL-1,$ and for this to be possible it becomes necessary to have 
\begin{equation}\label{lplnp}
\boxed{L_PL\leq N_P}\cdot 
\end{equation}

\subsection{FDKD Pilot Patterns in DS-OFDM Systems}
All pilot symbols in a FDKD pilot cluster, except the middle one, are equal to zero. Furthermore, the non-zero pilot symbols of all pilot clusters are assumed to be the same and equal to $\chi$. It then follows that ${\bf P}_{\rm pat}(:,(L_P-1)/2)=\left[\chi, \ldots,\chi\right]^T$. All other columns of ${\bf P}_{\rm pat}$ comprise of entries equal to 0. It follows from \eqref{pandq}, \eqref{manyeq} and \eqref{spanofl} that 
\begin{equation}\label{fdkdfirsteq}
{\mathcal L}\left( {{\bm\theta} _l } \right) = {\mathcal L}\left( {{\bf f}^{\left( l \right)} } \right)\cdot
\end{equation}
This ensures that the second condition in \eqref{conditions} and  \eqref{tetalhtetal} are satisfied. However, a key requirement for this to be possible is that ${\bf f}^{\left( m \right)H} {\bf f}^{\left( n \right)}  = 0, 0\leq m,n \leq L-1$. This implies that we must have
\begin{equation}\label{npl}
\boxed{N_P\geq L}\cdot
\end{equation}
Furthermore, note that $r\left({\bm \theta}_l\right)=1,\:\:r\left({\bar{\bm \theta}}_l\right) = 2B_c+1$. If the BMC and RNC-BEM are satisfied, the first condition in \eqref{conditions} is satisfied if
\[
2B_c+1\geq Q\cdot
\]
More discussion on the above equation is presented in Sec. V.

\subsection{DS-MIMO-OFDM Systems}
{\it The system model and  the channel estimation matrix, $\boldsymbol {\mathcal P}$, is an extension of the one derived in \cite{pilot_JSP_07} and differs from the one derived in \cite{comments} in two aspects}, namely, {\it i}) the channel estimation scheme in \cite{comments} considered the case of $O=1$ only, while the schemes in this paper and \cite{pilot_JSP_07} consider $O>1$, {\it ii}) as per the scheme of \cite[(6), (7)]{comments}, a demodulated pilot symbol belonging to the $i$th pilot cluster takes into account the contributions of pilot symbols only in the $i$th pilot cluster, whereas, the schemes presented in this paper \eqref{ypb} and \cite{pilot_JSP_07} take into account the effect of all pilot clusters.

Without loss of generality, we assume that all quantities correspond to the zeroth MIMO receiver. Furthermore, we assume that there are $N_T$ transmit antennas. Let ${\bf h}_l^{(i)}$ denote a vector of BEM parameters associated with the $l$th path between Transmitter $i$ and Receiver 0. Define ${\bf h}^{(i)} = \left[{\bf h}_0^{(i)}, \ldots, {\bf h}_{L-1}^{(i)}\right]^T$ as the concatenation of all ${\bf h}_l^{(i)}, l=0,\ldots, L-1$.  The quantity ${\bf P}_{\rm pat}^{(i)}$ is defined in a way similar to that in \eqref{pandq} and denotes the pilot-pattern matrix used by Transmitter $i$. Replacing ${\bf P}_{\rm pat}$ in \eqref{manyeq} by ${\bf P}_{\rm pat}^{(i)}$, one obtains ${\underline{\bm\theta}}_l^{(i)}$. Using ${\underline{\bm\theta}}_l^{(i)}$ in \eqref{manyeq}, one obtains ${\bf R}_l^{\left( k, i \right)}  = {\underline{\bm\theta}} _l^{(i)} \left( {k:N_p  + k - 1} \right)$, ${{\bm\theta}} _l^{(i)}  = \left[ {{\bf R}_l^{\left( 0,i \right)} ,}  {{\bf R}_l^{\left( 1,i \right)} ,}  { \ldots ,}  {{\bf R}_l^{\left( {N_P  - 1},i \right)} }\right]$. The channel estimation matrix associated between Transmitter $i$ and Receiver 0 is denoted as ${\boldsymbol{\mathcal P}}^{(i)}$ where 
\begin{equation}\label{pi}
\begin{array}{*{20}c}
 {\boldsymbol{\mathcal P}}^{(i)} = \left[ {\begin{array}{*{20}c}
   {\underbrace {\begin{array}{*{20}c}
   {{\bm \theta} _0^{(i)} {\bf W}^{\left( { - B_c } \right)} {\bf B}}  \\
    \vdots   \\
   {{\bm \theta} _0^{(i)} {\bf W}^{\left( {B_c } \right)} {\bf B}}  \\
\end{array}}_{{\bm \phi} _0^{(i)} }} & {\begin{array}{*{20}c}
    \ldots   \\
    \vdots   \\
    \ldots   \\
\end{array}} & {\underbrace {\begin{array}{*{20}c}
   {{\bm \theta} _{L - 1}^{(i)} {\bf W}^{\left( { - B_c } \right)} {\bf B}}  \\
    \vdots   \\
   {{\bm \theta} _{L - 1}^{(i)} {\bf W}^{\left( {B_c } \right)} {\bf B}}  \\
\end{array}}_{{\bm \phi} _{L - 1}^{(i)} }}  \\
\end{array}} \right], & {\bm \phi} _l^{(i)}  = \underbrace{\left[ {\begin{array}{*{20}c}
   {{\bm \theta} _l ^{(i)}} & {} & {}  \\
   {} &  \ddots  & {}  \\
   {} & {} & {{\bm \theta} _l^{(i)} }  \\
\end{array}} \right]}_{{\bar {\bm \theta}}_l^{(i)}}\underbrace {\left[ {\begin{array}{*{20}c}
   {{\bf W}^{\left( { - B_c } \right)} {\bf B}}  \\
    \vdots   \\
   {{\bf W}^{\left( {B_c } \right)} {\bf B}}  \\
\end{array}} \right]}_{\bf E}   \cdot
\end{array}
\end{equation}The quantities ${\underline{\bm\theta}}_l^{(i)}$, ${\bf R}_l^{\left( k, i \right)}$, ${{\bm\theta}} _l^{(i)}$ and ${{\bm \phi} _l^{(i)} }$ are associated with the $l$th path between Transmitter $i$ and Receiver 0.
The system model \eqref{yeqhp} can be rewritten as 
\begin{equation}\label{yeqhpmimo}
{\bf \bar y} = {\boldsymbol{\mathcal P}}{\bf h}+{\bf n}_1
\end{equation}
where ${\boldsymbol{\mathcal P}}=\left[{\boldsymbol{\mathcal P}}^{(0)}, \ldots, {\boldsymbol{\mathcal P}}^{(N_T-1)}\right]$, ${\bf h} = \left[{\bf h}^{(0)T}, \ldots, {\bf h}^{(N_T-1)T}\right]^T$ and ${\bf n}_1$ is the interference due to data clusters and noise. For the case of DS-MIMO-OFDM systems, \eqref{conditions} can be rewritten as 
\begin{equation}\label{mimoconditions}
\begin{array}{*{20}l}
r\left({\bm \phi}_l^{(i)}\right) &=& Q,&0\leq l\leq L-1,\:\:0\leq i\leq N_T-1\\{\mathcal{L}}\left({\bm \phi}_{l_1}^{(i)}\right) &\bot& {\mathcal{L}}\left({\bm \phi}_{l_2}^{(j)}\right),& (i,l_1)\neq (j, l_2) \cdot
\end{array}
\end{equation} Similar to \eqref{tetalhtetal}, ${\bf P}_{\rm pat}^{(i)}, 0\leq i \leq N_T-1,$ should be designed such that
\begin{equation}\label{mimotetalhtetal}
{\bm\theta}_{l_1}^{(i)H}{\bm\theta}_{l_2}^{(j)} = 0, \:\:\:\: (i,l_1)\neq (j, l_2)\cdot
\end{equation}
When the BEMC is satisfied and none of the basis vectors of $\bf E$ lies in the null space of ${{\bar {\bm \theta}}_l^{(i)}}, 0\leq l \leq L-1, 0\leq i \leq N_T-1,$ the first condition in \eqref{mimoconditions} is satisfied. If we design ${\bf P}_{\rm pat}^{(i)}$ such that 
\begin{equation}\label{ppatdsofdmmimo}
\boxed{{\bf P}_{\rm pat}^{(i)} = \left[{\bf f}^{(iL_PL)},{\bf f}^{(iL_PL+L)}, \ldots, {\bf f}^{(iL_PL+(L_P-1)L)}\right]}
\end{equation}
it follows from the definitions of ${\bf R}_l^{\left( k, i \right)}$ and ${{\bm\theta}} _l^{(i)}$ that
\begin{equation}
\begin{array}{*{20}c}\label{feq}
   {{\mathcal L}\left( {{\bm\theta} _l^{(i)} } \right) = {\mathcal L}\left( {{\bf R}_l^{\left( 0,i \right)} } \right) =  \cdots  = {\mathcal L}\left( {{\bf R}_l^{\left( {N_P  - 1},i \right)} } \right) = {\mathcal L}\left( {\bf f}^{(iL_PL+l)},{\bf f}^{(iL_PL+L+l)}, \ldots, {\bf f}^{(iL_PL+(L_P-1)L+l)} \right)},  \\
   {\bm\theta}_{l_1}^{(i)H}{\bm\theta}_{l_2}^{(j)} = 0, \:\:\:\: (i,l_1)\neq (j,l_2)  
\end{array}
\end{equation}
which ensures that ${\boldsymbol{\mathcal P}}$ is of full column rank. Similar to the analysis that was done for DS-OFDM systems, it is necessary to have
\begin{equation}\label{lplntnp}
\boxed{L_PLN_T\leq N_P}\cdot
\end{equation}

The FDKD pilot pattern has only one non-zero pilot symbol at the center of the pilot cluster, i.e., the only non-zero column in ${\bf P}_{\rm pat}^{(i)}$ is the $(L_P-1)/2$st column. If we design ${\bf P}_{\rm pat}^{(i)}$ such that 
\begin{equation}\label{ppatdsofdmmimofdkd}
\boxed{{\bf P}_{\rm pat}^{(i)}\left(:,\frac{L_P-1}{2}\right) = {\bf f}^{(iL)}}
\end{equation}
it follows from the definitions of ${\bf R}_l^{\left( k, i \right)}$ and ${{\bm\theta}} _l^{(i)}$ that
\begin{equation}
\begin{array}{*{20}c}
   {{\mathcal L}\left( {{\bm\theta} _l^{(i)} } \right) = {\mathcal L}\left( {{\bf R}_l^{\left( 0,i \right)} } \right) =  \cdots  = {\mathcal L}\left( {{\bf R}_l^{\left( {N_P  - 1},i \right)} } \right) = {\mathcal L}\left( {\bf f}^{(iL+l)} \right)},  \\
   {\bm\theta}_{l_1}^{(i)H}{\bm\theta}_{l_2}^{(j)} = 0, \:\:\:\: (i,l_1)\neq (j,l_2)  
\end{array}
\end{equation}
which ensures that ${\boldsymbol{\mathcal P}}$ is of full column rank. Similar to the analysis that was done for DS-OFDM systems, it is necessary to have
\begin{equation}\label{lntnp}
\boxed{LN_T\leq N_P}\cdot
\end{equation}

\section{Miscellaneous Conditions for DS-OFDM Systems and consensus with prior art}
We now derive some additional conditions on the pilot cluster length, BEM order, number of pilot clusters that need to be satisfied for the full column-rank condition. Additionally, these are found to be in agreement with the ones derived in \cite{commentscomments} and \cite{pilot_SPAWC_07}.

It follows from \eqref{lthetalprev} that 
\[
r\left( {{\bm\theta} _l } \right)=r\left( {{\bf R}_l^{\left( 0 \right)} } \right) =  \cdots  = r\left( {{\bf R}_l^{\left( {N_P  - 1} \right)} } \right)= L_P\cdot
\]
Note that ${{\bf R}_l^{\left( 0 \right)} }, \ldots, {{\bf R}_l^{\left( {N_P  - 1} \right)} },$ all have $L_P$ columns. However, if the FDKD pilot pattern is used, there is only one non-zero column (the FDKD pilot pattern has only one non-zero pilot symbol per pilot cluster) and consequently, we have
\[
r\left( {{\bm\theta} _l } \right)=r\left( {{\bf R}_l^{\left( 0 \right)} } \right) =  \cdots  = r\left( {{\bf R}_l^{\left( {N_P  - 1} \right)} } \right)= 1\cdot
\]
It therefore follows (depending on the pilot-pattern design) that
\begin{equation}
L_P\geq r\left( {{\bm\theta} _l } \right)\geq 1\cdot
\end{equation}

Consider \eqref{phil}. It follows that $r\left( {{\bar{\bm\theta}} _l } \right) = (2B_c+1)r\left( {{\bm\theta} _l } \right)$. Consequently,
\begin{equation}\label{rankrange}
L_P(2B_c+1)\geq r\left( {{\bar{\bm\theta}} _l } \right)\geq (2B_c+1)\cdot
\end{equation}
It follows from \eqref{conditions} and \eqref{ranke} that $r\left({\bm \phi}_i\right)=Q$ and $r\left(\bf E\right)=Q$. Since $r\left({\bf AB}\right)\leq{\rm min}\{r\left(\bf A\right), r\left(\bf B\right)\}$, it follows from \eqref{phil} that $r\left( {{\bar{\bm\theta}} _l } \right)\geq Q$. From \eqref{rankrange}, it therefore follows that
\begin{equation}\label{firstproof}
r\left( {{\bar{\bm\theta}} _l } \right)\left(2B_c+1\right)\geq Q\cdot
\end{equation}
We now show how some of the results derived in this paper agree with those in \cite{commentscomments} and \cite{pilot_SPAWC_07}.
\begin{itemize}
	\item {\it Consensus with \cite{commentscomments} }: The length of the observation cluster used in \cite{commentscomments} is $O=2B_c+1=1$, or in other words, $B_c=0$. From \eqref{firstproof} it follows that
\begin{equation}
\boxed{L_P\geq Q}\cdot
\end{equation}	
This will be called the pilot-cluster BEM condition (PC-BEM). {\it Note that PC-BEM is similar to {\it Lemma 1} derived in \cite{commentscomments}}. Though this is derived for a DS-OFDM system, a similar derivation can be done for the DS-MIMO-OFDM system.
\item {\it Consensus with \cite{pilot_SPAWC_07}} : Assumption 1 in \cite{pilot_SPAWC_07} states that all subcarriers are used for channel estimation (all subcarriers are pilot subcarriers). Assumption 2 in \cite{pilot_SPAWC_07} states that the pilot cluster is a FDKD pilot cluster. It therefore follows that $N=N_PL_P, L_P=2B_c+1, G=0$. Since FDKD pilot patterns are used, it follows from \eqref{npl} that $N_P\geq L$. This satisfies a part of the condition stated in \cite[(18)]{pilot_SPAWC_07}.

It follows from \eqref{fdkdfirsteq} that $r\left({\bm \theta}_l\right)=1$. From the definition of ${\bar{\bm \theta}}_l$ in \eqref{phil}, it follows that $r\left({\bar{\bm \theta}}_l\right) = 2B_c+1 = L_P$. Consequently, \eqref{firstproof} translates to $L_P\geq Q$. Since $N=N_PL_P$, we have $N/N_P\geq Q $. Together with the previous condition, we now have
\begin{equation}\label{secondproof}
\boxed{\frac{N}{Q}\geq N_P \geq L}
\end{equation}
which completely agrees with  \cite[(18)]{pilot_SPAWC_07}.

\end{itemize}
\section{Simulation Studies}
\begin{table}[t]
	\centering
	\caption{Parameter Sets used in Simulation Studies.}\label{dimensions}
		\begin{tabular}{p{1.2cm}|@{}p{13.9cm}  }
			\hline\hline
\parbox{.07\textwidth}{\centering {\bf Parameter Set}}& \parbox{.63\textwidth}{\centering {\bf Parameter Set Values}}\\ \hline\hline
\parbox{.07\textwidth}{\centering $S_1$}&\parbox{.63\textwidth}{\centering $\left\{N=128, Q=3, P_b=1, L_P=3, B_c=1, P_{\rm sep}=8,  L=4, f_D=0.1, L_PL=12, N_P=16, QL=12\right\}$ }\\ \hline
\parbox{.07\textwidth}{\centering $S_2$}&\parbox{.63\textwidth}{\centering $\left\{N=256, Q=3, P_b=1, L_P=3, B_c=1, P_{\rm sep}=16,  L=4, f_D=0.1, L_PL=12, N_P=16, QL=12\right\}$ }\\ \hline
\parbox{.07\textwidth}{\centering $S_3$}&\parbox{.63\textwidth}{\centering $\left\{N=512, Q=3, P_b=1, L_P=3, B_c=1, P_{\rm sep}=16,  L=4, f_D=0.1, L_PL=12, N_P=32, QL=12\right\}$ }\\ \hline
\parbox{.07\textwidth}{\centering $S_4$}&\parbox{.63\textwidth}{\centering $\left\{N=1024, Q=5, P_b=2, L_P=5, B_c=2, P_{\rm sep}=16,  L=4, f_D=0.3, L_PL=20, N_P=64, QL=20\right\}$ }\\ \hline
\hline
\end{tabular}
\end{table}

\begin{table}[t]
	\centering
	\caption{Summary of the Scenarios for which the Full Column-Rank Conditions are tested.}\label{remarks}
		\begin{tabular}{p{1.2cm}|@{}p{13.9cm}  }
			\hline\hline
\parbox{.07\textwidth}{\centering {\bf Parameter Set}}& \parbox{.93\textwidth}{\centering {\bf Remarks}}\\ \hline\hline
\parbox{.07\textwidth}{\centering $S_1$}&\parbox{.88\textwidth}{Satisfies $L_PL \leq N_P$ \eqref{lplnp}. Pilot patterns designed as per \eqref{ppatdsofdm} and Sec. IV.B. ensure full column-rank of $\boldsymbol{\mathcal{P}}$ in DS-OFDM and FDKD-based DS-OFDM systems, respectively. }\\ \hline
\parbox{.07\textwidth}{\centering $S_2$}&\parbox{.88\textwidth}{Satisfies $L_PL \leq N_P$ \eqref{lplnp}. Pilot patterns designed as per \eqref{ppatdsofdm} and Sec. IV.B. ensure full column-rank of $\boldsymbol{\mathcal{P}}$ in DS-OFDM and FDKD-based DS-OFDM systems, respectively. }\\ \hline
\parbox{.07\textwidth}{\centering $S_3$}&\parbox{.88\textwidth}{ 
\begin{itemize}
	\item Satisfies $L_PL \leq N_P$ \eqref{lplnp}. Pilot patterns designed as per \eqref{ppatdsofdm} and Sec. IV.B. ensure full column-rank of $\boldsymbol{\mathcal{P}}$ in DS-OFDM and FDKD-based DS-OFDM systems, respectively.
	\item Satisfies  $L_PLN_T\leq  N_P$ \eqref{lplntnp}. Pilot patterns designed as per \eqref{ppatdsofdmmimo} ensure full column-rank of $\boldsymbol{\mathcal{P}}$ in a DS-MIMO-OFDM system with $N_T=2$ transmitters.
	\item Satisfies  $LN_T\leq  N_P$ \eqref{lntnp}. Pilot patterns designed as per \eqref{ppatdsofdmmimofdkd} ensure full column-rank of $\boldsymbol{\mathcal{P}}$ in a FDKD-based DS-MIMO-OFDM system with $N_T=2, \ldots,6$ transmitters.
\end{itemize}
 }\\ \hline
\parbox{.07\textwidth}{\centering $S_4$}&\parbox{.88\textwidth}{\begin{itemize}
	\item Satisfies $L_PL \leq N_P$ \eqref{lplnp}. Pilot patterns designed as per \eqref{ppatdsofdm} and Sec. IV.B. ensure full column-rank of $\boldsymbol{\mathcal{P}}$ in DS-OFDM and FDKD-based DS-OFDM systems, respectively.
	\item Satisfies  $L_PLN_T\leq  N_P$ \eqref{lplntnp}. Pilot patterns designed as per \eqref{ppatdsofdmmimo} ensure full column-rank of $\boldsymbol{\mathcal{P}}$ in a DS-MIMO-OFDM system with $N_T=2,3$ transmitters.
	\item Satisfies  $LN_T\leq  N_P$ \eqref{lntnp}. Pilot patterns designed as per \eqref{ppatdsofdmmimofdkd} ensure full column-rank of $\boldsymbol{\mathcal{P}}$ in a FDKD-based DS-MIMO-OFDM system with $N_T=2, \ldots,6$ transmitters.
\end{itemize} }\\ \hline
\hline
\end{tabular}
\end{table}

Four BEM models are considered in the simulation studies, namely, P-BEM, CE-BEM, S-BEM and GCE-BEM. Full column-rank condition of the channel estimation matrix, ${\boldsymbol{\mathcal P}}$, is tested in various scenarios. Though not explicitly mentioned, it will be understood that all the four BEM models will be tested in each scenario. All the four BEM models satisfy the BEMC for each of the parameter sets in Table \ref{dimensions}. The RNC-BEM condition is tested for the parameter sets in Table \ref{dimensions} in the context of  DS-OFDM, FDKD-based DS-OFDM, DS-MIMO-OFDM and FDKD-based DS-MIMO-OFDM systems. In each case, it is found that all the BEM models satisfy the RNC-BEM condition.

The full column-rank condition of the channel estimation matrix, ${\boldsymbol{\mathcal P}}$, is tested for the parameter sets in Table \ref{dimensions} in the context of a DS-OFDM system and FDKD-based DS-OFDM system. When the pilot-pattern matrix, ${\bf P}_{\rm pat}^{(i)}$, is designed according to \eqref{ppatdsofdm} and Sec. III.B, it ensures the full column-rank condition of ${\boldsymbol{\mathcal P}}$ in DS-OFDM and FDKD-based DS-OFDM systems, respectively. Similarly, for the case of DS-MIMO-OFDM system, Parameter set $S_3$ is tested with two transmitters ($N_T=2$), while Parameter set $S_4$ is tested with two and three transmitters ($N_T=2,3$). It is found that the pilot-pattern matrix, ${\bf P}_{\rm pat}^{(i)}$, designed as per \eqref{ppatdsofdmmimo}, ensures the full column-rank condition of ${\boldsymbol{\mathcal P}}$ in a DS-MIMO-OFDM system. Furthermore, for the case of FDKD-based DS-MIMO-OFDM system, Parameter sets $S_3, S_4$ are tested with two to six transmitters ($N_T=2, \ldots, 6$). It is found that the pilot-pattern matrix, ${\bf P}_{\rm pat}^{(i)}$, designed as per \eqref{ppatdsofdmmimofdkd}, ensures the full column-rank condition of ${\boldsymbol{\mathcal P}}$ in a FDKD-based DS-MIMO-OFDM system.

The summary of the various scenarios are given in Table \ref{remarks}.

\section{Conclusions}
We derived and discussed the conditions for the full column-rank condition of the channel estimation matrix in a DS-OFDM system. Firstly, it was shown that a BEM model needs to satisfy a BEMC. Secondly, the BEM model and the pilot-pattern design need to satisfy the RNC-BEM condition. The RNC-BEM condition was derived in the context of a DS-OFDM system. Once the BEM and RNC-BEM conditions were satisfied, it was shown how to design the pilot patterns (the pilot-pattern matrix) that ensure the channel estimation matrix is of full column rank in a DS-OFDM system. As special cases, we considered the FDKD pilot pattern. The pilot-pattern design method was extended to the case of a DS-MIMO-OFDM system with as many as six transmitters. The PC-BEM condition, which relates the pilot cluster length to the BEM order, was derived for a DS-MIMO-OFDM system and was shown to be similar to that published in a recent journal. The relationships between the number of subcarriers, number of pilot clusters, BEM order and multipath length was derived and was shown to be similar to that derived in a recent conference paper.

\section*{Acknowledgement}
We thank STMIcroelectronics Asia pacific Pte. Ltd. for sponsorship of this research activity. The first author thanks Denny H. Leung, department of mathematics, National university of Singapore, for preliminary discussions on the rank-nullity theorem. We thank reviewers of an earlier submission (T-SP-09319-2009) for their useful comments which improved the quality of this manuscript in many ways. Section V was added due to Reviewer 1's comments. The use of different pilot clusters was a result of the comments of Reviewer 2, while the physical interpretations and some derivations in Sec. IV. A were due to the comments of Reviewer 3. 
\section{Appendix RNC-BEM}
The RNC-BEM condition will be derived only for DS-OFDM systems. For other cases like DS-MIMO-OFDM systems, one can derive it in a similar way. The purpose of the derivation is to show that the example of pilot-pattern design and BEM models considered in this paper indeed satisfy the RNC-BEM condition. {\it Hence, we mathematically derive the relationship between the pilot patterns and BEM matrix for the sake of completeness though one can easily verify the RNC-BEM condition in Matlab or a computer-aided approach}.

Let ${\bf e}_0, \ldots, {\bf e}_{Q-1},$ denote the basis vectors of $\bf E$. The $Q$ orthonormal bases that span the column space of $\bf B$ are denoted by ${\bf b}_0, \ldots, {\bf b}_{Q-1}$. Furthermore, let \footnote{If ${\bf b}_q$ lies in the null space of ${\bf W}^{(i)}$, then ${\bf w}^{(i,q)}={\bf 0}$. If ${\bf b}_q$ does not lie in the null space of ${\bf W}^{(i)}$, then from the rank-nullity theorem, ${\bf w}^{(i,q)}$ is a basis vector of ${\bf W}^{(i)}{\bf B}$. Note that if the BEMC is satisfied there is at least one non-zero ${\bf w}^{(i,q)}, i=-B_c,\ldots, B_c$.}
\begin{equation}\label{wiq}
{\bf w}^{(i,q)}={\bf W}^{(i)}{\bf b}_q, q=0,\ldots, Q-1\cdot
\end{equation}
From \eqref{phil} and the rank-nullity theorem, we have the relationship 
\begin{equation}\label{eq}
{\bf e}_q=\left[{\bf w}^{(-B_c,q)T},\ldots, {\bf w}^{(B_c,q)T}\right]^T\cdot
\end{equation} The RNC-BEM condition states that none of the vectors ${\bf e}_0, \ldots, {\bf e}_{Q-1},$ should lie in the null space of ${{\bar {\bm \theta}}_l}$, i.e., 
\[
{{\bar {\bm \theta}}_l}{\bf e}_Q \neq {\bf 0}, q=0,\ldots,Q-1,\:\:\:\:l=0,\ldots,L-1\cdot
\]
In what follows, we derive the condition for a given $l$ and $q$. From \eqref{phil} and \eqref{eq}, it follows that ${{\bar {\bm \theta}}_l}{\bf e}_Q \neq {\bf 0}$ implies that at least one of ${\bm \theta}_l{\bf w}^{(i,q)}, i=-B_c,\ldots, B_c,$ should be a non-zero vector. Assuming that ${\bf w}^{(i,q)}\neq {\bf 0}$, we analyze the conditions when ${\bm \theta}_l{\bf w}^{(i,q)}\neq {\bf 0}$. 

Recall from \eqref{lthetalprev} that when ${\bf P}_{\rm pat}$ is designed as in \eqref{ppatdsofdm}, we have 
\begin{equation}\label{rl0}
{\mathcal L}\left( {{\bf R}_l^{\left( 0 \right)} } \right) =  \cdots  = {\mathcal L}\left( {{\bf R}_l^{\left( {N_P  - 1} \right)} } \right)\cdot
\end{equation} Furthermore, the various columns of ${{\bf R}_l^{\left( 0 \right)} }$ are orthogonal to each other, i.e., ${{\bf R}_l^{\left( 0 \right)H} }{{\bf R}_l^{\left( 0 \right)} }=L_P{\bf I}_{L_P}$. Define an index vector ${\bm \rho}_j$ as ${\bm \rho}_j=\left[j, j+L_p, \ldots, (N_P-1)L_P+j\right], j=0,\ldots, L_P-1$. 

\begin{figure}[t]
     \centering 
      \vspace {0.0in} 
        \includegraphics [width = 6cm,height = 4cm] {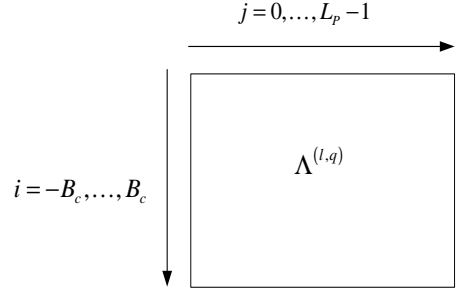}
         \caption{ Structure of the Matrix $\Lambda^{\left(l,q\right)}$.}\label{appendix}
         \vspace{0.0in} 
\end{figure}

We can write 
\begin{equation}
{\bm \theta}_l{\bf w}^{(i,q)}=\sum_{j=0}^{j=L_P-1}{\bm \theta}_l(:,{\bm \rho}_j){\bf w}^{(i,q)}({\bm \rho}_j)\cdot
\end{equation}
Recall from Fig. \ref{matrix1}, that all columns of ${\bm \theta}_l(:,{\bm \rho}_j)$ span the same column space.
From the definitions in \eqref{manyeq}, \eqref{lthetalprev}, \eqref{rl0} and Fig. \ref{matrix1}, it follows that
\begin{equation}
\mathcal{L}\left({\bm \theta}_l(:,\rho_j(0))\right)=\ldots=\mathcal{L}\left({\bm \theta}_l(:,\rho_j(N_P-1))\right)={\mathcal L}\left({\bf f}^{(jL+l)}\right),\:\:\: 0\leq j\leq L_P-1\cdot
\end{equation}
 The choice of ${\bf P}_{\rm pat}$ as in \eqref{ppatdsofdm} ensures that 
 \begin{equation}
 {\mathcal L}\left({\bm \theta}_l(:,{\bm \rho}_j)\right)={\mathcal L}\left({\bf f}^{(jL+l)}\right)
 \end{equation} which implies that 
 \begin{equation}
 {\mathcal L}\left({\bm \theta}_l(:,{\bm \rho}_j)\right)\bot{\mathcal L}\left({\bm \theta}_l(:,{\bm \rho}_k)\right), j\neq k
 \end{equation}
  i.e.,  
  \begin{equation}
  {\bm \theta}_l(:,{\bm \rho}_j)^H{\bm \theta}_l(:,{\bm \rho}_k)={\bf 0}, j\neq k\cdot
  \end{equation}
  This means that for ${\bm \theta}_l{\bf w}^{(i,q)}$ to be a non-zero vector, at least one of the summands in $\sum_{j=0}^{L_P-1}{\bm \theta}_l(:,{\bm \rho}_j){\bf w}^{(i,q)}({\bm \rho}_j)$ should be a non-zero vector. Let ${\bm \Lambda}^{(l,q)}$ be a matrix associated with ${{\bar {\bm \theta}}_l}$ and ${\bf e}_Q$. Furthermore, ${\bm \Lambda}^{(l,q)}(i,j)$ is equal to 1 if ${\bm \theta}_l(:,{\bm \rho}_j){\bf w}^{(i,q)}({\bm \rho}_j)$ is a non-zero vector and equal to 0 if ${\bm \theta}_l(:,{\bm \rho}_j){\bf w}^{(i,q)}({\bm \rho}_j)$ is a null vector. This is depicted in Fig. \ref{appendix}. 
The RNC-BEM condition can be now stated as 
\begin{equation}\label{rncbem}
{\bm \Lambda}^{(l,q)}\neq {\bf 0},\:\:\:\: l=0,\ldots, L-1,\:\:q=0,\ldots, Q-1\cdot
\end{equation}
Note that ${\bm \Lambda}^{(l,q)}\neq {\bf 0}$ implies that ${\bf e}_q$ does not lie in the null space of ${{\bar {\bm \theta}}_l}$. We now derive ${\bm \Lambda}^{(l,q)}(i,j)$ in terms of ${\bf b}_q$, the $q$th basis vector of the BEM matrix $\bf B$. For this we need to examine, in detail, the quantity ${\bf w}^{(i,q)}({\bm \rho}_j)$.

From \eqref{wiq}, it follows that ${\bf w}^{(i,q)}({\bm \rho}_j)={\bf W}^{(i)}({\bm \rho}_j,:){\bf b}_q$. Consider the $n$th column of ${\bf W}^{(i)}({\bm \rho}_j,:)$. From the definition of ${\bf W}^{(i)}$ in \eqref{manyeq}, it follows that 
\begin{equation}\label{one}
{\bf W}^{(i)}({\bm \rho}_j,n)= e^{+j2\pi\gamma^{(j,i)} \Delta fn}{\bf f}^{(\left\langle n\right\rangle_{N_p})*}
\end{equation} where $\gamma^{(j,i)} = {\bf p}_1(j)-P_b-i$.  Note that ${\bm \theta}_l(:,{\bm \rho}_j)$ is a symmetric (circulant) matrix and as stated before, 
\begin{equation}\label{lthetal}
{\mathcal L}\left({\bm \theta}_l(:,{\bm \rho}_j)\right)={\mathcal L}\left({\bf f}^{(jL+l)}\right)\cdot
\end{equation}
This implies that 
\begin{equation}\label{array}
\begin{array}{*{20}c}
{\bm \theta}_l(:,{\bm \rho}_j){\bf W}^{(i)}({\bm \rho}_j,n)&\neq& {\bf 0},&&&& \left\langle n\right\rangle_{N_p}&=&jL+l,\\
{\bm \theta}_l(:,{\bm \rho}_j){\bf W}^{(i)}({\bm \rho}_j,n)&=& {\bf 0},&&&& \left\langle n\right\rangle_{N_p}&\neq&jL+l\\\cdot
\end{array}
\end{equation}
Define ${\bf n}^{(j,l)}=\left[jL+l, jL+l+N_P, \ldots, jL+l+(P_{\rm sep}-1)N_P\right]$. From \eqref{one}, \eqref{lthetal} and \eqref{array}, it follows that 
\begin{equation}
{\bm \theta}_l(:,{\bm \rho}_j){\bf w}^{(i,q)}({\bm \rho}_j)={\bm \theta}_l(:,{\bm \rho}_j){\bf W}^{(i)}({\bm \rho}_j,:){\bf b}_q={\bm \theta}_l(:,{\bm \rho}_j){\bf f}^{(jL+l)*}\psi^{(l,q,i,j)}
\end{equation}
where $\psi^{(l,q,i,j)}=\sum_{k=0}^{P_{\rm sep}-1}{\bf b}_q\left({\bf n}^{(j,l)}(k)\right)e^{j2\pi\gamma^{(j,i)}\Delta f{\bf n}^{(j,l)}(k)}$. Since it follows from \eqref{lthetal} that ${\bm \theta}_l(:,{\bm \rho}_j){\bf f}^{(jL+l)*}\neq {\bf 0}$, ${\bm \Lambda}^{(l,q)}(i,j)$ is equal (not equal) to 0 if $\psi^{(l,q,i,j)}$ is equal (not equal) to 0. We therefore arrive at
\begin{equation}
\boxed{
\begin{array}{*{20}l}
{\bm \Lambda}^{(l,q)}(i,j) &=& 1-\delta\left(\psi^{(l,q,i,j)}\right), & {\bf w}^{(i,q)}\neq {\bf 0} \\
{\bm \Lambda}^{(l,q)}(i,j) &=& 0, & {\bf w}^{(i,q)} = {\bf 0}
\end{array}}\cdot
\end{equation}


\end{document}